# Influence of magnetic fields on the performance of spin-orbit torque magnetic random-access memory*

LIU Jiaxin, ZHOU Yuqing, SHI Guoyi, CAI Kaiming†

School of Physics, Huazhong University of Science and Technology, Wuhan 430074, China

† Corresponding author. E-mail: kmcai@hust.edu.cn

**Abstract**

Spin-orbit torque magnetic random-access memory (SOT-MRAM) possesses high speed, ultrahigh endurance, and excellent compatibility with advanced semiconductor manufacturing processes, and is considered to be a promising nonvolatile memory technology. However, the free layer in the magnetic tunnel junction (MTJ) is affected by an intrinsic bias field ($H_s$) originating from the stray field of the reference layer and interlayer coupling associated with surface roughness. The bias field gives rise to a pronounced asymmetry in the critical current density for magnetization switching between the two resistance states, thereby increasing the overall energy consumption. Recent solutions typically introduce additional magnetic layers within the MTJ stack to compensate for $H_s$. However, such an approach increases manufacturing costs and limits its practicality in wafer-scale manufacturing. To address the issue of asymmetry, we propose a method that avoids modifying the original MTJ stack. The basic idea is to regulate the write current via local stray magnetic field engineering, which involves filling magnetic materials into designated vertical interconnect access (VIA) channels during the back-end-of-line (BEOL) process. For a representative SOT-MTJ with perpendicular magnetic anisotropy (PMA), in which the undesired $H_s$ is typically directed along the z-axis, micromagnetic simulations show that a ferromagnetic filling layer magnetized in plane can significantly reduce write-current asymmetry and provide the auxiliary field required for deterministic switching. Furthermore, by slightly displacing the MTJ from the center, the bias-compensation effect can be further optimized, reducing the write-current bias ratio from 21.6% in the conventional design to 1.3%. Notably, this approach enables field-free switching—a critical feature for SOT-MTJ applications targeting high integration density. The concept is also applicable for SOT-MTJs with in-plane magnetic anisotropy. Finally, the scaling analysis indicates excellent compatibility: even when scaled to 20% of the original size (MTJ diameter ≈ 10 nm), the write-current bias ratio remains at a low level of 0.1%, indicating that our design is effective over a broad range of MTJ dimensions and highly suitable for high-density integration with advanced technology nodes.

**Keywords:** spin-orbit torque; magnetic random-access memory; symmetric switching current; advanced technology node

**DOI:** 10.7498/aps.75.20251720

**CSTR:** 32037.14.aps.75.20251720

**Publication record:** Acta Physica Sinica, Vol. 75, No. 6 (2026), 060807 · SPECIAL TOPIC—Applied magnetism · COVER ARTICLE · Received 15 December 2025; revised manuscript received 22 January 2026 · Project supported by the National Natural Science Foundation of China (Grant No. 12404133).

## 1 Introduction

The rapid advancement of big data and artificial intelligence technologies has imposed higher demands on the speed, power consumption, and capacity of memory within information processing systems. Ideal memory technology must possess key characteristics such as ultrahigh speed, low power consumption, high stability, and high integration density. Additionally, it requires nonvolatility to meet the requirements of emerging application scenarios[1,2]. However, mainstream charge-based memory technologies are currently constrained by their intrinsic physical mechanisms. These technologies struggle to achieve an effective balance among core performance metrics such as speed, power consumption, and reliability. This limitation has become a primary bottleneck restricting system performance improvements[3,4]. To overcome the limitations of traditional memory, magnetic random-access memory (MRAM) has attracted significant attention due to its high-speed operation, high endurance, and nonvolatility. In particular, SOT-MRAM, which is based on the spin-orbit torque (SOT) effect, offers write speeds at the sub-nanosecond level (<1 ns). It also demonstrates excellent read/write endurance (>$10^{15}$ cycles). Furthermore, its good compatibility with advanced semiconductor processes makes it widely regarded as a strong candidate for next-generation nonvolatile memory technology[5-7].

---

* The paper is an English translated version of the original Chinese paper published in Acta Physica Sinica. Please cite the paper as: J. Liu, Y. Zhou, G. Shi, and K. Cai, Influence of magnetic fields on the performance of spin-orbit torque magnetic random-access memory, Acta Physica Sinica 75, 060807 (2026). doi: 10.7498/aps.75.20251720.

A typical SOT-MRAM memory cell adopts a three-terminal structure with separate read and write paths, as shown in Figure 1. For writing, the bottom heavy-metal (HM) layer generates a spin current. Reading relies on the tunneling magnetoresistance (TMR) effect within the top magnetic tunnel junction (MTJ)[8]. Based on the orientation of the magnetocrystalline anisotropy in the free layer (FL) of the MTJ, SOT-MRAM can be categorized into two main types. One type has perpendicular magnetic anisotropy (PMA). In this configuration, the magnetization easy axis of the free layer is perpendicular to the layer interfaces, and the MTJ adopts a cylindrical design. Because PMA is governed by interfacial material properties, it supports high integration density. The other type has in-plane magnetic anisotropy (IMA). Here, the magnetization easy axis is parallel to the interfaces. Its magnetic anisotropy is primarily determined by shape anisotropy, and the MTJ is typically designed with an elliptical shape. Moreover, the write performance of the device (such as critical current and switching speed) is closely related to the angle between the direction of the magnetization easy axis and the current direction[9,10].

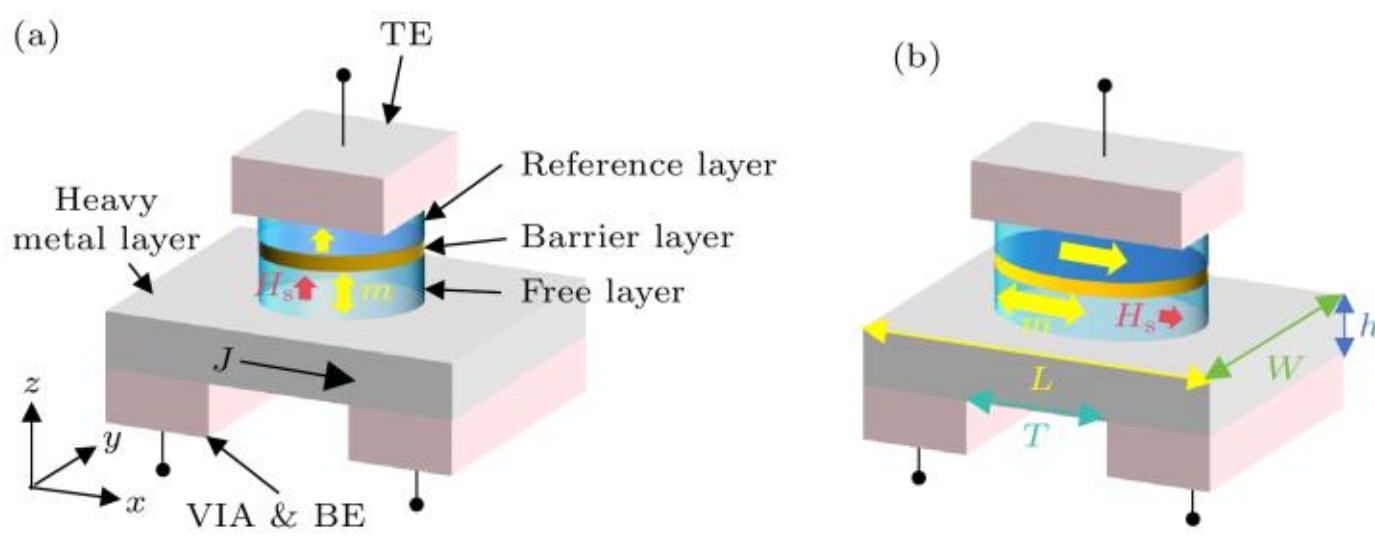

**Fig. 1.** Schematic of an SOT-MRAM cell: (a) PMA structure, where TE and BE denote the top and bottom electrodes, respectively; (b) IMA structure.

However, both PMA and IMA devices share a common issue: stray fields generated by the reference layer (RL) and interlayer coupling induced by interface roughness create an intrinsic bias field at the free layer[11]. This bias field not only causes significant asymmetry in the write current and external magnetic field required for switching between different resistance states but also introduces a series of engineering challenges. For instance, the transition requiring the larger write current dissipates more power and may reduce reliability[12]. The corresponding drive transistor size must increase, thereby reducing integration density. Furthermore, the bias field reduces thermal stability and causes device uniformity issues, which increases the complexity of external circuit design and calibration[13,14]. These factors collectively limit the performance improvement of SOT-MRAM. Although these phenomena are widely reported[15,16], discussions and solutions regarding these issues remain scarce[17,18]. Existing solutions typically rely on introducing additional magnetic layers within the MTJ (such as synthetic antiferromagnetic structures or other auxiliary stack layers) to cancel the bias field. While such methods can achieve compensation at the device level, they require precise control of exchange coupling and film thickness uniformity across multiple interfaces at the nanoscale, resulting in an extremely narrow process window. In wafer-level manufacturing, minor fluctuations in thickness or interfaces lead to compensation mismatch, thereby amplifying the dispersion of write current asymmetry within and between chips. Additionally, increasing the number of thin-film layers significantly raises process complexity, testing, and calibration costs, while compromising array yield and scalability[19,20]. Therefore, a bias field compensation method that does not modify the MTJ thin films and is compatible with existing processes is needed to support the development of SOT-MRAM toward higher performance and density.

Drawing on the magnetic-field modulation of the SOT-MRAM write current, we propose a new SOT-MRAM cell architecture that is highly compatible with back-end-of-line (BEOL) integration. At a prescribed depth in the vertical interconnect access (VIA), part of the conventional nonmagnetic metal fill is replaced by a ferromagnetic material directly beneath the SOT channel. The resulting localized stray field modulates the write current. We first use micromagnetic simulations to systematically examine how uniform magnetic fields affect the write current. We then characterize the nonuniform stray field generated by the proposed architecture and evaluate its ability to reduce the write current and suppress write-current bias. Scaling simulations show that the architecture retains effective bias compensation after substantial dimensional reduction, indicating compatibility with advanced semiconductor technology nodes. This scheme therefore offers a promising route to improve the energy efficiency, uniformity, and integration density of SOT-MRAM.

## 2 Effects of Write-Current Bias and Magnetic-Field Modulation

### 2.1 Effects of Write-Current Bias

Studies indicate that the switching behavior of SOT-MRAM depends on the external magnetic field[21-23]. Ideally, due to the rotational symmetry of the system, its bipolar switching should exhibit highly symmetric characteristics. However, in practical devices, stray fields from the reference layer and interlayer coupling induced by interface roughness inevitably affect the free layer. These effects are equivalent to applying an effective bias magnetic field $H_s$ to the free layer (as shown in Figure 1), causing asymmetry in the switching behavior between different resistance states[11,24-26]. Physically, this equivalent bias field is not determined by a single factor. For example, equivalent magnetic charges formed at the pattern edges of the reference layer generate a dipolar stray field at the free layer, which usually provides the main static bias component. Interface roughness may also induce additional interlayer coupling fields. Furthermore, the distribution of write current in the heavy metal channel and interconnect structures introduces an Oersted field component at the free layer. Since these contributions have different sensitivities to material stacking, pattern size, and process variations, the $H_s$ of each cell in actual wafer arrays often varies. This leads to a discrete distribution of write current bias and energy barrier asymmetry at the array scale, increasing the difficulty of unified writing and calibration by peripheral circuits. As shown in Figure 2(a), the switching fields on the left and right sides of the hysteresis loop are $H_{cl}$ and $H_{cr}$, respectively. In the presence of the bias field $H_s$, the coercive fields on both sides are no longer symmetric, i.e., $|H_{\mathrm{cl}}| \neq |H_{\mathrm{cr}}|$. The coercive field of the free layer is typically defined as $H_{\mathrm{c}} = |H_{\mathrm{cl}} - H_{\mathrm{cr}}|/2$, and the bias magnetic field can be expressed as $H_{\mathrm{s}} = |H_{\mathrm{cl}} + H_{\mathrm{cr}}|/2$. Similarly, the write current density changes from the ideal case of $|J_{AP\to P}| = |J_{P\to AP}| = J_0$ to $|J_{AP\to P}| = J_0 - \Delta J_1$ and $|J_{P\to AP}| = J_0 + \Delta J_2$. Here, $J_{AP\to P}$ is the critical current density for switching from the antiparallel (AP) state to the parallel (P) state, and $J_{P\to AP}$ is the critical current density for switching from the P state to the AP state (the magnetization direction of the reference layer follows Figure 1(a) and (b)). $J_0$ is the critical write current density without the influence of $H_s$, and $\Delta J_1$ and $\Delta J_2$ are the current density bias amounts caused by the bias field (as shown in Figure 2(b)).

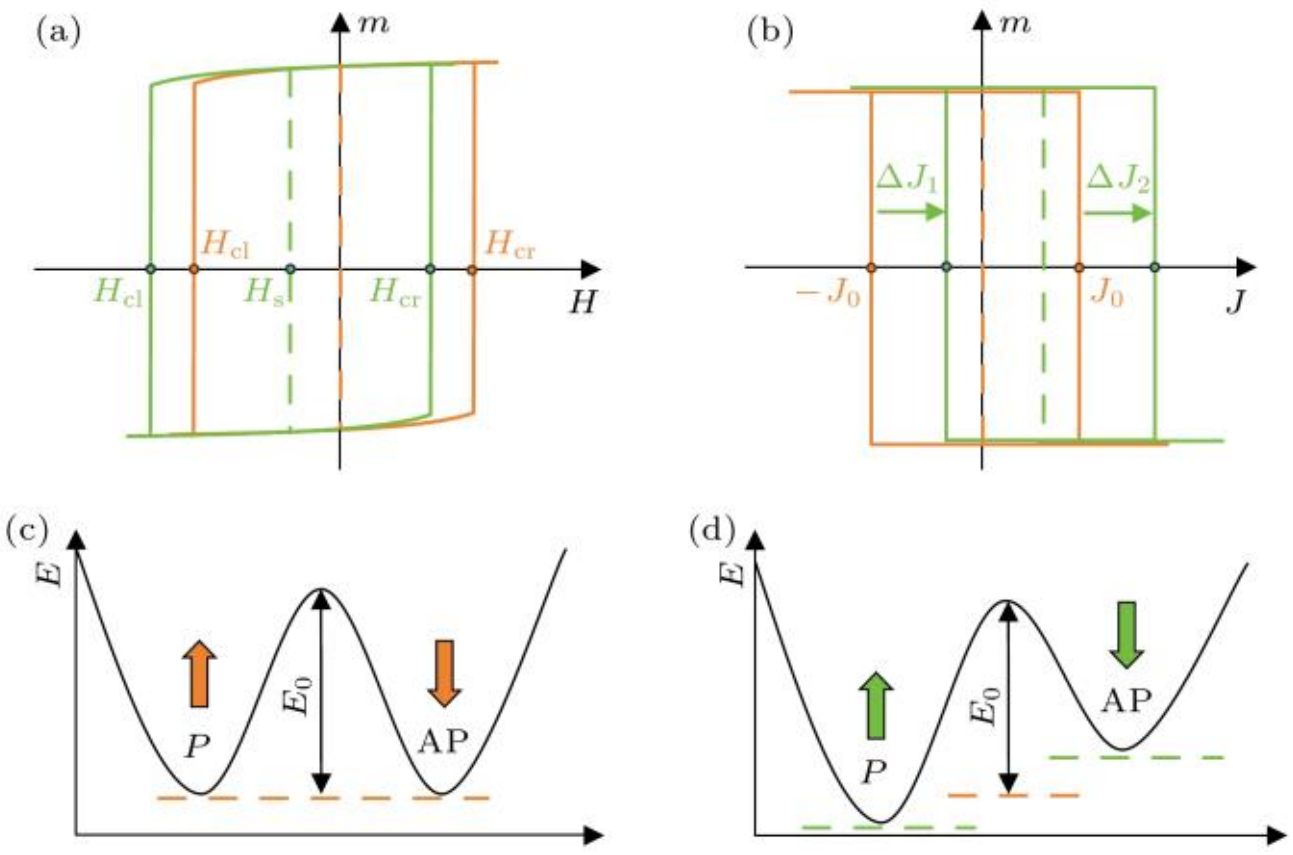

**Fig. 2.** Schematic illustration of how a bias magnetic field affects SOT-MRAM performance: (a) shift of the magnetic hysteresis loop and asymmetry of the coercive fields; (b) asymmetric critical switching current densities; (c) energy barriers without a bias field; (d) unequal energy barriers of the P and AP states in the presence of a bias field.

As shown in Figure 1, the heavy metal layer is characterized by its thickness h, width W, length L, and resistivity ρ. In the absence of a bias field, the total power consumption $P_0$ for completing a full AP→P→AP switching cycle is given by

$$P_0 = 2I_0^2R = 2\rho hWLJ_0^2, \tag{1}$$

If a bias field exists and approximately satisfies $\Delta J_1 = \Delta J_2 = \Delta J$ within a certain range, the write power consumption $P_s$ is given by

$$P_\mathrm{s} = \rho hWL[(J_0 - \Delta J)^2 + (J_0 + \Delta J)^2], \tag{2}$$

Consequently, the minimum fractional increase in power consumption caused by the current bias is $\Delta P/P_0 = \Delta J^2/J_0^2$, where $\Delta P = 2\rho hWL\Delta J^2$. Thus, reducing ΔJ directly contributes to lowering power consumption.

However, implementing separate write operations with distinct $J_{AP\to P}$ and $J_{P\to AP}$ values in practical circuits requires dual current sources or programmable current driver structures[27-29]. This approach increases control logic complexity and peripheral circuit area, particularly because each memory cell or sub-array in large-scale arrays needs an independent current regulation module[14,26,30]. Furthermore, switching current paths introduces control delays from address decoding and multiplexer switching, along with the settling time required for current source stabilization. These delays limit the write throughput of the memory during high-speed operations[14,31]. Therefore, to ensure integration density and reduce write latency, practical SOT-MRAM designs employ a unified write current to simplify circuit modules. In this scenario, the power consumption $P_s$ for completing one AP→P→AP operation is at least $P_\mathrm{s} = 2\rho hWL(J_0 + \Delta J)^2$, resulting in additional power consumption of $\Delta P = 2\rho hWL(2J_0\Delta J + \Delta J^2)$. This extra power dissipation is significant in high-density arrays, creating an urgent need to suppress the write-current bias caused by the bias field.

Additionally, the bias field affects the data-retention stability of SOT-MRAM. According to the Stoner-Wohlfarth model, the switching energy barrier $E_{P(AP)}$ from P (AP) to AP (P) can be described as[13]

$$E_{\mathrm{P(AP)}} = E_0\left(1 \pm \frac{H_\mathrm{s}}{H_\mathrm{K}^\mathrm{eff}}\right)^2, \tag{3}$$

Here, $E_0$ represents the energy barrier in the absence of a magnetic field, and $H_\mathrm{K}^\mathrm{eff}$ denotes the effective anisotropy field. As illustrated in Figures 2(c) and 2(d), when $H_s > 0$, the P state exhibits greater stability than the AP state. Consequently, the data retention time of the AP state decreases, thereby compromising data reliability.

### 2.2 Effects of Magnetic Fields on the Write Current in PMA and IMA Structures

Because the bias field $H_s$ can be treated as a component of the effective field $H_{eff}$ acting on the free layer, we analyze how magnetic fields modulate SOT-driven magnetization switching. The magnetization dynamics are described by the Landau-Lifshitz-Gilbert (LLG) equation incorporating the SOT term[32]:

$$\partial \boldsymbol{m}/\partial t = -\gamma\mu_0(\boldsymbol{m}\times \boldsymbol{H}_{eff}) + \alpha(\boldsymbol{m}\times \partial \boldsymbol{m}/\partial t) + \gamma\mu_0 H_{SOT}^{DL}[\boldsymbol{m}\times(\boldsymbol{\sigma}\times \boldsymbol{m})] + \gamma\mu_0 H_{SOT}^{FL}(\boldsymbol{m}\times\boldsymbol{\sigma}), \quad (4)$$

Here, m = M/$M_s$ denotes the reduced magnetization, where m = $m_x e_x + m_y e_y + m_z e_z$. The terms $m_x$, $m_y$, and $m_z$ represent the components of m along the x, y, and z directions, respectively. $\boldsymbol{\sigma}$ indicates the spin polarization direction, α is the damping constant, γ is the gyromagnetic ratio, and $\mu_0$ is the vacuum permeability. The SOT contribution comprises a damping-like torque (DLT) and a field-like torque (FLT). Their magnitudes are defined by $H_{SOT}^{DL} = \frac{\hbar\theta_{SH}J_{SOT}}{2eM_s t_F}$ and $H_{SOT}^{FL} = \eta H_{SOT}^{DL}$, respectively. In these expressions, e represents the elementary charge, $\theta_{SH}$ denotes the spin Hall angle, $J_{SOT}$ is the drive current density, $M_s$ is the saturation magnetization, ħ is the reduced Planck constant, $t_F$ is the free layer thickness, and η is the ratio of $H_{SOT}^{FL}$ to $H_{SOT}^{DL}$.

To analyze the influence of external magnetic fields on the SOT-driven magnetization switching process, we conducted systematic micromagnetic simulations under various magnetic field conditions[33]. Table 1 lists the key material and geometric parameters used. The material and geometric parameters in Table 1 are representative values selected from the typical ranges reported in Ref. [33] and are kept fixed throughout this study. Using one common parameter set allows the relative effects of the magnetic filling structure on the bias field, critical switching current, and write power consumption to be compared directly. The aim is to identify trends and physical mechanisms, not to determine an absolute optimum for a specific material system. When materials or process nodes change, relevant critical values will vary with parameters such as $M_s$, $K_u$, and $\theta_{SH}$. However, the relative trends identified here still serve as a reference for structural design. Figures 3(a) and 3(b) illustrate the device schematics for the PMA and IMA structures, respectively. Specifically, the free layer of the PMA structure is modeled as a cylinder with a diameter of 50 nm. The free layer of the IMA structure is modeled as an elliptical cylinder with a major axis of 50 nm and a minor axis of 25 nm. Given the significant shape anisotropy of the IMA free layer, we examined three typical geometric configurations. In the first configuration, the easy axis of the free layer is parallel to the write current direction (i.e., φ = 0°). In the second configuration, the easy axis forms an intermediate angle with the write-current direction, 0° < φ < 90° (with φ = 30° taken as an example). Honjo et al.[9] fabricated and tested actual devices at this angle. Their work provides guidance for selecting $H_s$ and helps align simulations with practical scenarios. In the third configuration, the easy axis is perpendicular to the write current direction (i.e., φ = 90°). Figure 3(c) displays the switching dynamics of the free layers in different structures. These include the dependence on external magnetic fields and write speed.

**Table 1.** Parameters used in the micromagnetic simulations.

| **Parameter** | **PMA** | **IMA** |
|---|---|---|
| Saturation magnetization, $M_s$ (A m$^{-1}$) | $0.9 \times 10^6$ | $1.1 \times 10^6$ |
| Uniaxial anisotropy constant, $K_u$ (J m$^{-3}$) | $550.0 \times 10^3$ | $38.2 \times 10^3$ |
| Exchange stiffness constant, $A_{ex}$ (J m$^{-1}$) | $1.2 \times 10^{-11}$ | $1.3 \times 10^{-11}$ |
| Spin Hall angle, $\theta_{SH}$ | 0.1 | 0.1 |
| Ratio of field-like to damping-like SOT effective fields, $\eta$ | −0.3 | 0.5 |
| Dzyaloshinskii-Moriya interaction strength (J m$^{-2}$) | $1.3 \times 10^{-3}$ | $0.2 \times 10^{-3}$ |
| Damping constant, $\alpha$ | 0.02 | 0.1 |
| Free-layer thickness, $t_F$ (nm) | 1.0 | 1.5 |
| Current-pulse width, $\tau$ (ns) | 1.0 | 1.0 (5.0)* |

* The value in parentheses (5.0 ns) is used only for the IMA device with $\varphi$ = 90°.

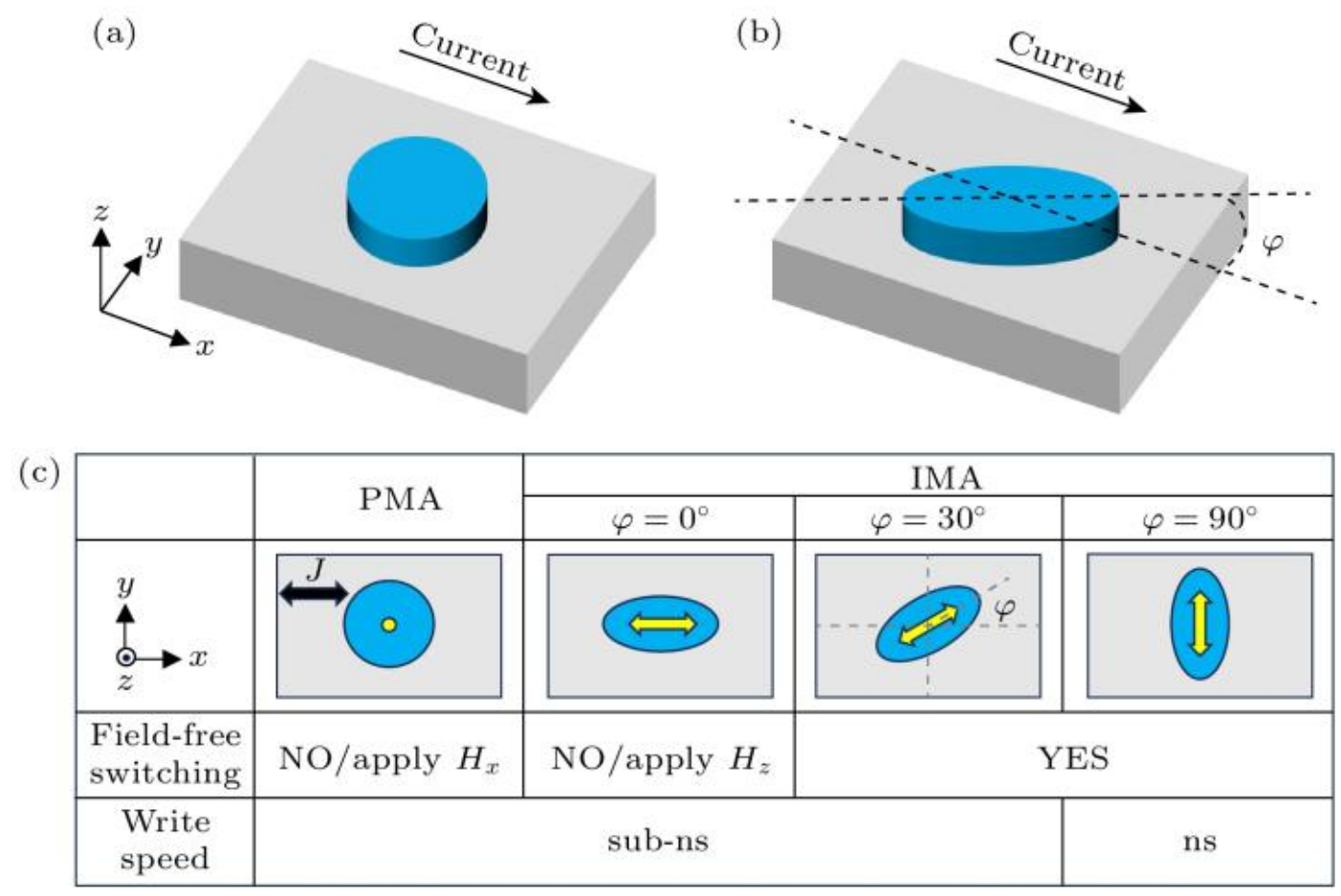

**Fig. 3.** Micromagnetic simulation geometries: (a) PMA structure; (b) IMA structure; (c) switching dynamics of the PMA and IMA structures. The gray and blue regions represent the heavy-metal layer and the free layer, respectively.

Figure 4 presents the simulation results for the PMA structure. To achieve deterministic switching of the free layer driven by current in the PMA structure, an auxiliary field $H_x$ along the x-direction is typically required. As shown in Figure 4(a), when the external magnetic field is small, the critical switching current density $J_c$ decreases linearly with increasing $H_x$, and the direction of $H_x$ determines the switching polarity. Furthermore, when a magnetic field $H_y$ along the y-direction is introduced under a fixed auxiliary field of $\mu_0H_x$ = 60 mT, the modulation of $J_c$ by $H_y$ exhibits nonlinear characteristics, as illustrated in Figure 4(b). In contrast, under the same fixed auxiliary field condition of $\mu_0H_x$ = 60 mT, the modulation of $J_c$ by the magnetic field $H_z$ along the z-direction varies linearly, as shown in Figure 4(c).

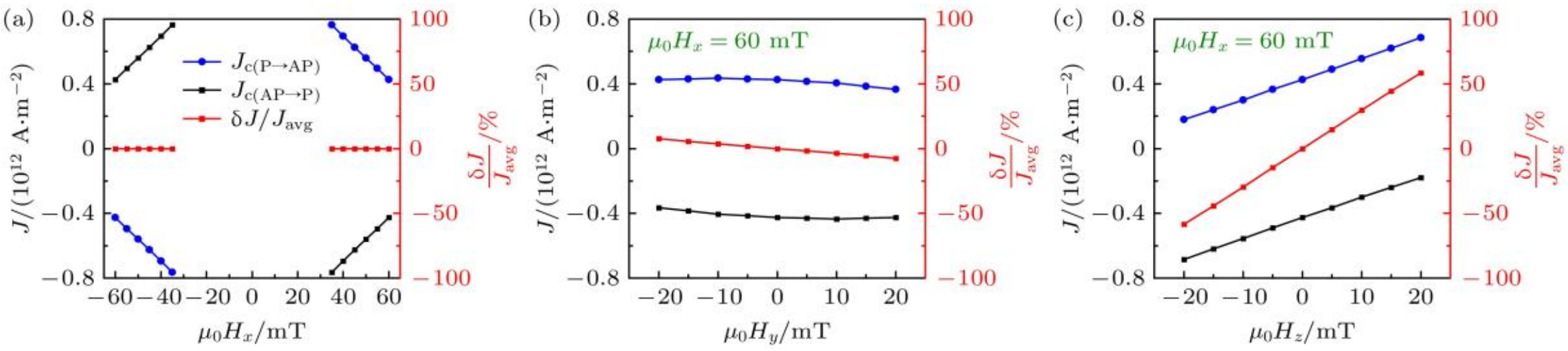

**Fig. 4.** Modulation of the critical switching current density and bias ratio in the PMA structure under external uniform magnetic fields: (a) Dependence of the critical switching current density and bias ratio on $H_x$; (b) dependence of the critical switching current density and bias ratio on $H_y$ with $\mu_0 H_x = 60$ mT; (c) dependence of the critical switching current density and bias ratio on $H_z$ with $\mu_0 H_x = 60$ mT.

To quantitatively characterize the magnitude of the critical switching-current bias, we introduce the bias ratio $\delta J/J_{avg}$, defined as follows:

$$\frac{\delta J}{J_{avg}} = \frac{|J_{c(P\to AP)}| - |J_{c(AP\to P)}|}{|J_{c(P\to AP)}| + |J_{c(AP\to P)}|},$$

Here, $\delta J = (|J_{c(P\to AP)}|-|J_{c(AP\to P)}|)/2$ represents the bias current density, and $J_{avg} = (|J_{c(P\to AP)}|+|J_{c(AP\to P)}|)/2$ denotes the average critical switching current density. Based on this definition, Figures 4(a)-(c) illustrate the variation of the bias ratio under different external magnetic field conditions. Simulation results indicate that under the sole influence of $H_x$, $J_c$ exhibits a symmetric linear decay as $|H_x|$ increases. Specifically, $|J_{c(P\to AP)}| = |J_{c(AP\to P)}|$, so no $J_c$ bias is introduced; this agrees with the findings of Taniguchi et al.[23]. However, with a fixed $H_x > 0$ serving as an auxiliary field (where $J_{c(P\to AP)} > 0$ and $J_{c(AP\to P)} < 0$), $H_y > 0$ results in a negative bias ratio $\delta J/J_{avg} < 0$, whereas $H_z > 0$ yields a positive bias ratio $\delta J/J_{avg} > 0$. This demonstrates that $H_y$ and $H_z$ exert opposite biasing effects on $J_c$ under the same switching polarity. Leveraging the biasing effects of $H_y$ and $H_z$ can effectively compensate for the write-current bias. At identical magnetic field strengths, the magnitude of the current bias induced by $H_z$ is significantly greater than that induced by $H_y$. This implies that $H_z$ dominates the influence on current bias in PMA structures.

Similarly, relevant simulations were conducted for IMA devices, with results presented in Figure 5. Compared to PMA structures, IMA structures exhibit switching dynamics that depend not only on the external magnetic field vector but also on the angle φ of the magnetization easy axis, due to the presence of in-plane shape anisotropy.

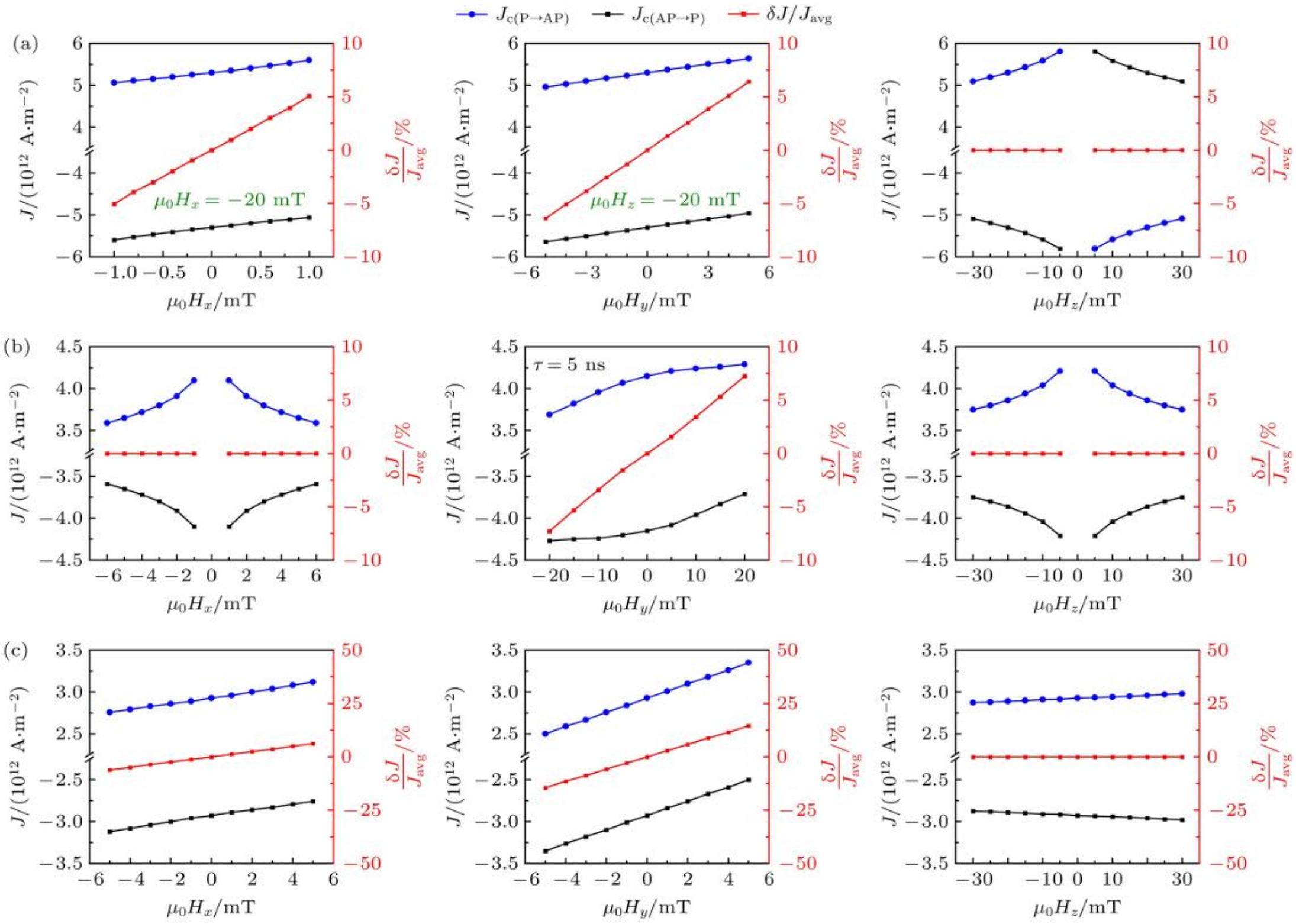

**Fig. 5.** Effect of an external magnetic field on the critical switching current density and bias ratio of the IMA structure with different easy-axis angles (φ): (a) Dependence of $J_c$ and $\delta J/J_{avg}$ on $H_x$, $H_y$, and $H_z$ for φ = 0°; (b) the corresponding curves for φ = 90°; (c) the corresponding curves for φ = 30° (note: an auxiliary field of $\mu_0H_z = -20$ mT is applied during the $H_x$ and $H_y$ scans at φ = 0° to assist deterministic magnetization switching; for the $H_y$ scans at φ = 90°, a current pulse width of τ = 5 ns is used).

For φ = 0°, as shown in Figure 5(a), a constant auxiliary field $\mu_0H_z = -20$ mT was applied in the simulation to break the out-of-plane symmetry and achieve deterministic switching. Under these conditions, the introduction of an in-plane magnetic field $H_x$ or $H_y$ breaks the symmetry of the switching barrier. This leads to a significant linear variation of the critical switching current density $J_c$ with the magnetic field strength. Specifically, as $H_x$ or $H_y$ is swept from negative to positive values, the bias ratio $\delta J/J_{avg}$ changes monotonically and linearly from negative to positive. This implies that the sign of the write current bias ratio can be flexibly adjusted by changing the direction of the in-plane magnetic field. In other words, the relative magnitudes of $|J_{c(P\to AP)}|$ and $|J_{c(AP\to P)}|$ can be tuned. In contrast, when the in-plane field is fixed at zero and only $H_z$ is varied (right panel of Figure 5(a)), $J_c$ decreases symmetrically and nonlinearly as $|H_z|$ increases. Throughout this process, the bias ratio $\delta J/J_{avg}$ remains zero regardless of the direction of $H_z$. This indicates that varying $H_z$ alone symmetrically modulates the system's switching barrier without introducing additional write asymmetry.

When φ = 90°, the field dependence differs markedly from that in the preceding case, as shown in Fig. 5(b). Varying $H_x$ along the hard axis causes a symmetric, nonlinear change in $J_c$. The bias ratio $\delta J/J_{avg}$ remains zero throughout the $H_x$ scanning range. This behavior suggests that $H_x$ along the hard axis acts similarly to the out-of-plane $H_z$. It symmetrically modulates the barrier height without inducing bias. However, the modulation of $J_c$ by $H_y$ along the easy axis exhibits significant nonlinear characteristics. Notably, when switching is driven solely by $H_y$ in this angular configuration, the dynamic process involves an incubation delay similar to that observed in spin-transfer torque (STT)-driven magnetization switching. Therefore, a long pulse with τ = 5 ns was used in the simulation. Under these conditions, the bias ratio $\delta J/J_{avg}$ varies nonlinearly with $H_y$, and its sign is determined by the magnetic field direction. These results indicate that the mechanism governing magnetic field modulation of the bias ratio undergoes a fundamental change when the easy axis is orthogonal to the current.

We further examined the intermediate angle $\varphi = 30°$ between the two aforementioned cases, as shown in Figure 5(c). At this angle, both the easy and hard axes form nonzero angles with the x and y axes. Consequently, the external magnetic fields $H_x$ and $H_y$ have components along both the easy and hard axes. As a result, both fields modulate $J_c$ linearly, as at $\varphi = 0°$. Specifically, the bias ratio $\delta J/J_{avg}$ shows a monotonic linear relationship with changes in $H_x$ or $H_y$, and its sign changes with the magnetic field direction. In contrast, the modulation of $J_c$ by $H_z$ retains its symmetric characteristic and does not introduce bias. This demonstrates that at non-orthogonal angles, magnetic field components in any in-plane direction can effectively break switching symmetry, thereby introducing a write current bias.

## 3 Micromagnetic Simulation of a New SOT-MRAM Architecture with Magnetic Filling

The preceding micromagnetic results show that a magnetic field applied in an appropriate direction can control and suppress write-current bias, thereby improving device symmetry. Practical high-density memory arrays, however, must generally operate without an external magnetic field. Moreover, process variations and spatial nonuniformity in the reference-layer stray field produce a distribution of equivalent bias fields across the array, making cell-level compensation necessary. Existing approaches insert additional magnetic functional layers into the MTJ stack. Although these layers can compensate the bias field, their performance depends on precise matching of layer thicknesses, interface roughness, and interlayer coupling; fabrication therefore requires coordinated control of multiple deposition and annealing steps, while the extra layers add process and test overhead. We therefore propose a SOT-MRAM cell architecture that is highly compatible with BEOL integration. A magnetic filling layer of prescribed thickness is inserted into the VIA, and its local stray field intrinsically compensates the bias field. Because the core functional layers of the MTJ stack remain unchanged, the design combines effective compensation with process feasibility and scalability to advanced technology nodes.

To evaluate the compensation effect, we used a conventional PMA structure with a representative bias as the baseline (hereafter, the 'conventional architecture'). Previous studies indicate that, in PMA MTJs, the ratio of the free-layer coercive field $H_c$ to the equivalent bias field $H_s$ from the reference-layer stray field and related effects is typically $H_c/H_s \approx 10/1$[34,35]. Using this ratio and the present simulation parameters, we set the equivalent bias field to $\mu_0 H_s = 15$ mT to represent a typical nonideal device environment [Fig. 6(a)]. Micromagnetic simulations show that when no bias field is present and an auxiliary field of $\mu_0 H_x = 30$ mT is applied, the critical write current densities are $|J_{c(P\rightarrow AP)}| = |J_{c(AP\rightarrow P)}| = 0.83\times10^{12}$ A/m$^2$. However, upon introducing a typical equivalent bias field of 15 mT, asymmetry appears in the write current: $|J_{c(P\rightarrow AP)}|$ increases to $1.01\times10^{12}$ A/m$^2$, while $|J_{c(AP\rightarrow P)}|$ decreases to $0.65\times10^{12}$ A/m$^2$, as shown in Figure 6(b). This indicates that the presence of the bias field significantly lowers the AP→P switching barrier and increases the P→AP switching barrier. The resulting bias ratio $\delta J/J_{avg}$ is 21.6%, which can substantially degrade device performance.

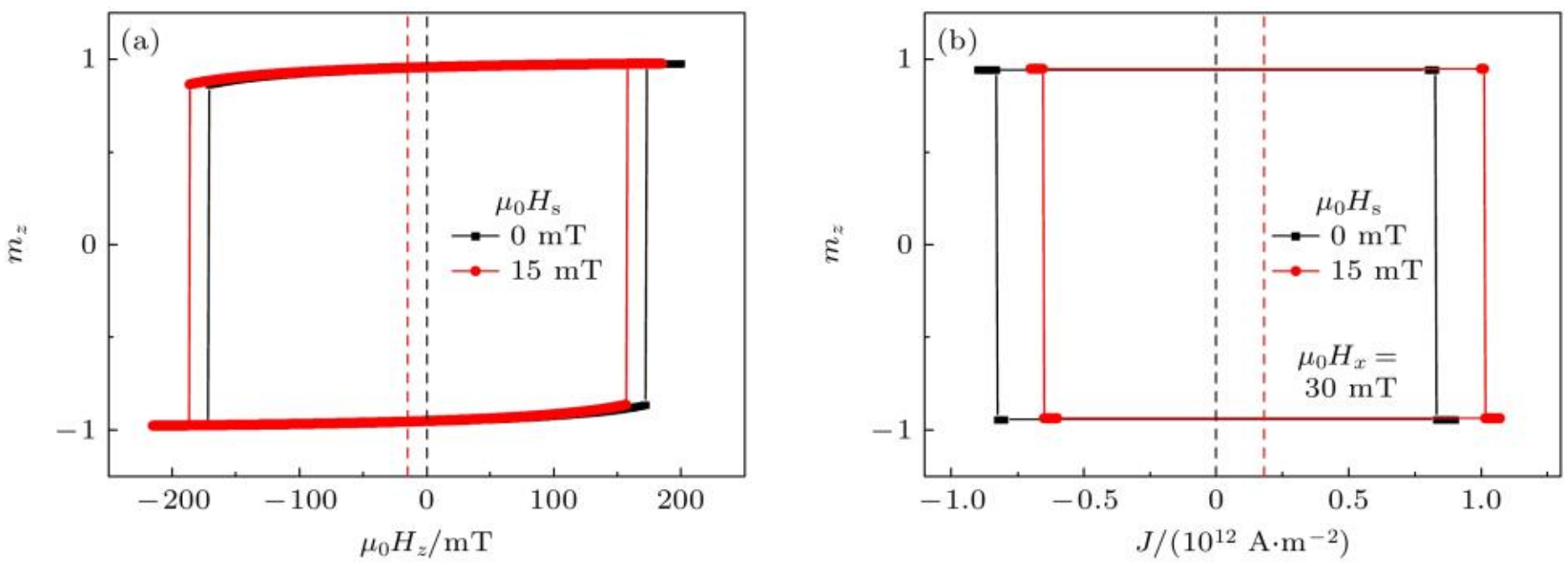

**Fig. 6.** Influence of bias field $\mu_0H_s$ = 15 mT: (a) Magnetic hysteresis loop of the free layer in the PMA structure under the influence of $\mu_0H_s$ = 15 mT; (b) $m_z$ - J curve of the free layer in the PMA structure under the influence of $\mu_0H_s$ = 15 mT with $\mu_0H_x$ = 30 mT.

To address this bias, we propose a cell architecture with CoFeB magnetic filling sections placed symmetrically in the VIAs at the two ends of the heavy-metal layer [Figs. 7(a) and 7(b)]. The key geometric parameters are set as follows[34,35]: the magnetic layer thickness H = 19 nm, with the magnetization direction along the +x axis; the spacing between the two magnetic filling sections, T = 150 nm; the length L = 420 nm and width W = 170 nm of the heavy metal layer in the memory cell; and the vertical distance h = 4 nm from the upper surface of the magnetic filling layer to the lower surface of the MTJ free layer (diameter d = 50 nm, layer thickness 1 nm). This design aims to construct a specific local magnetic field distribution within the plane where the MTJ free layer is located. Although a typical magnetic filling layer has a higher resistivity than a conventional VIA interconnect metal and therefore introduces additional power dissipation, the estimate in Appendix A indicates that this added dissipation is insufficient to offset the power savings from bias suppression.

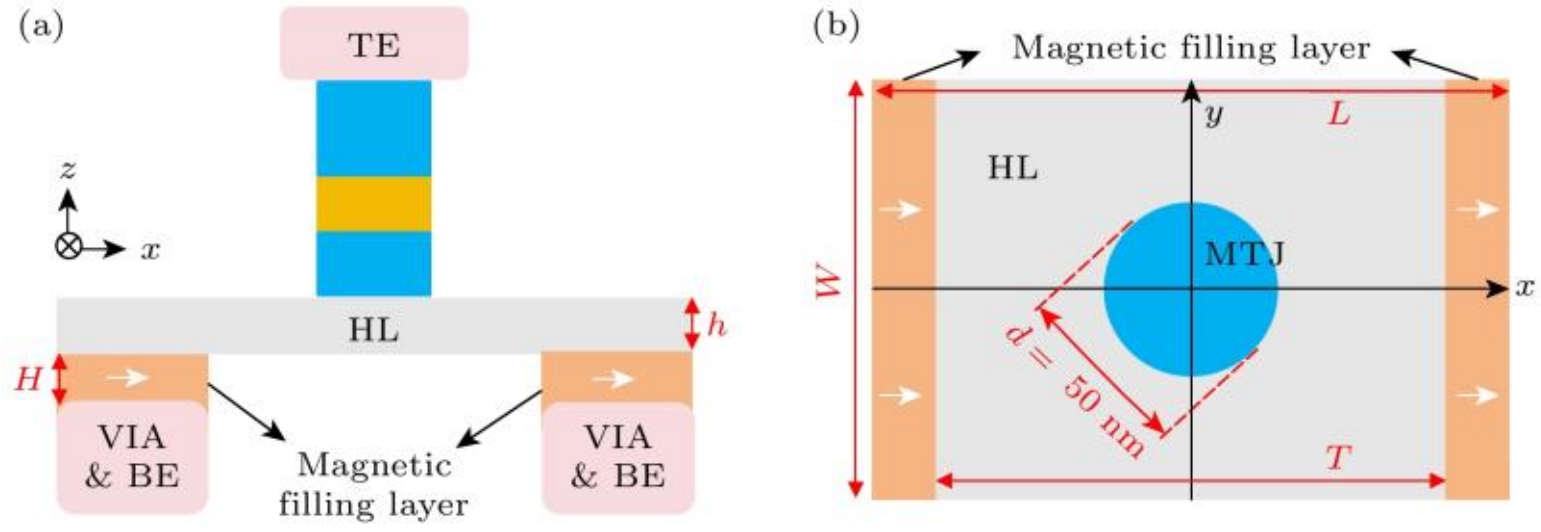

**Fig. 7.** Schematic of a PMA SOT-MRAM cell containing magnetic filling: (a) front view; (b) top view.

We analyzed the magnetic field distribution within the gray rectangular region of the free layer plane in Fig. 7(b), as illustrated in Fig. 8(a). A line scan of the magnetic field along Y = 0 nm (indicated by the green dashed line) yields the position-dependent magnetic-field components shown in Fig. 8(b). To optimize bias suppression, we performed systematic micromagnetic simulations of the PMA SOT-MRAM cell containing a magnetic filling layer and examined how the relative position of the MTJ and filling layer affects compensation of $H_s$. We established a coordinate system with the geometric center of the upper surface of the heavy-metal layer as the origin, as depicted in Fig. 7(b). We placed the MTJ center at nine representative positions: (±25, ±25), (±25, 0), (0, ±25), and (0, 0), denoted $P_1$-$P_9$, as shown in Fig. 8(a). The simulation results are presented in Fig. 8(c).

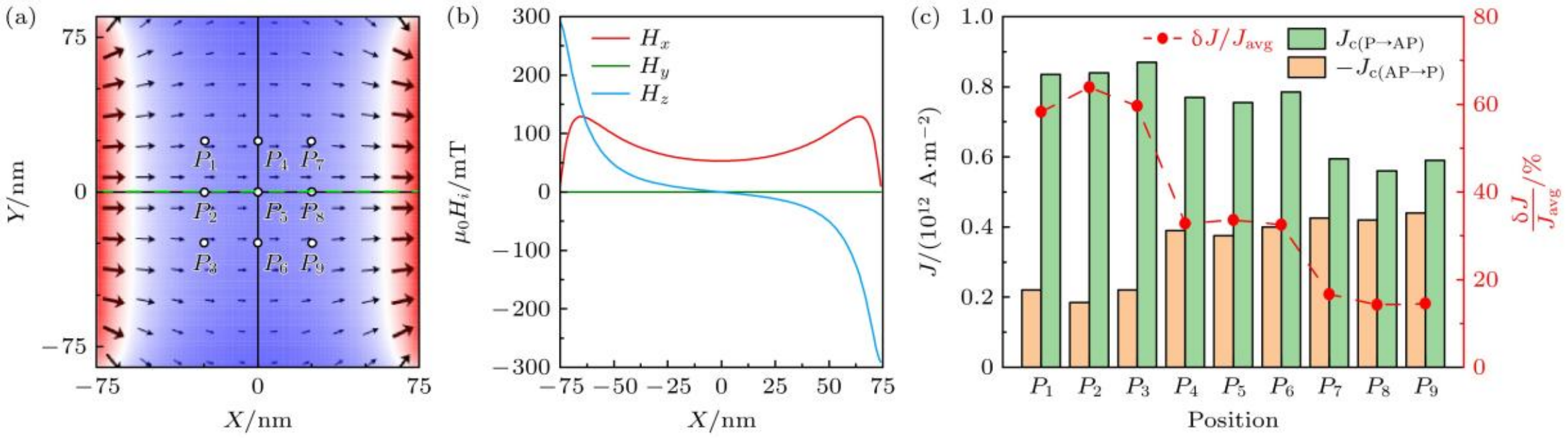

**Fig. 8.** Characterization of the local magnetic field distribution and the position-dependent compensation effect on the switching current bias: (a) Magnetic field distribution within the gray rectangular region in Fig. 7(b); (b) Distribution curves of magnetic field components in all directions along the X-axis at Y = 0 nm; (c) Critical switching current density and bias ratio at positions $P_1$-$P_9$.

First, the magnetic filling layer provides a magnetic field component in the +x direction. Consequently, the MTJ free layer can switch deterministically even without an external auxiliary magnetic field. Second, comparing the critical switching currents and bias ratios at different positions reveals that MTJs located in the positive x-axis region more effectively counteract the bias caused by $H_s$. The underlying physical mechanism is that the stray field generated by the magnetic filling layer in this region has a component in the −z direction. This component directly offsets the biasing effect of $H_s$. Meanwhile, the theoretical analysis in Section 2 indicates that the magnetic field component in the +y direction also helps mitigate the bias from $H_s$. Combining this with the magnetic field distribution in Fig. 8(a), the region satisfying both conditions (providing −z and +y components) lies in the fourth quadrant. Therefore, we conducted a refined search within the fourth quadrant. We selected the MTJ center coordinates (30, −52) as a typical optimized position, as shown in Fig. 9(a). Figure 9(b) presents the cross-sectional analysis of the magnetic field along Y = −52 nm. We compared the device with the proposed architecture at this position (Proposed) with the conventional device (Conventional). The results show that the bias ratio decreases significantly from 21.6% in the conventional architecture to 1.3%. This reduction essentially eliminates the negative impact of the equivalent bias field $H_s$, as illustrated in Fig. 9(c). Although producing such a large offset of the MTJ center may be difficult because of alignment tolerances and cell-size constraints, the analysis of these extreme positions reveals the theoretical optimization potential of this architecture. These findings provide important design guidance for optimizing the process window.

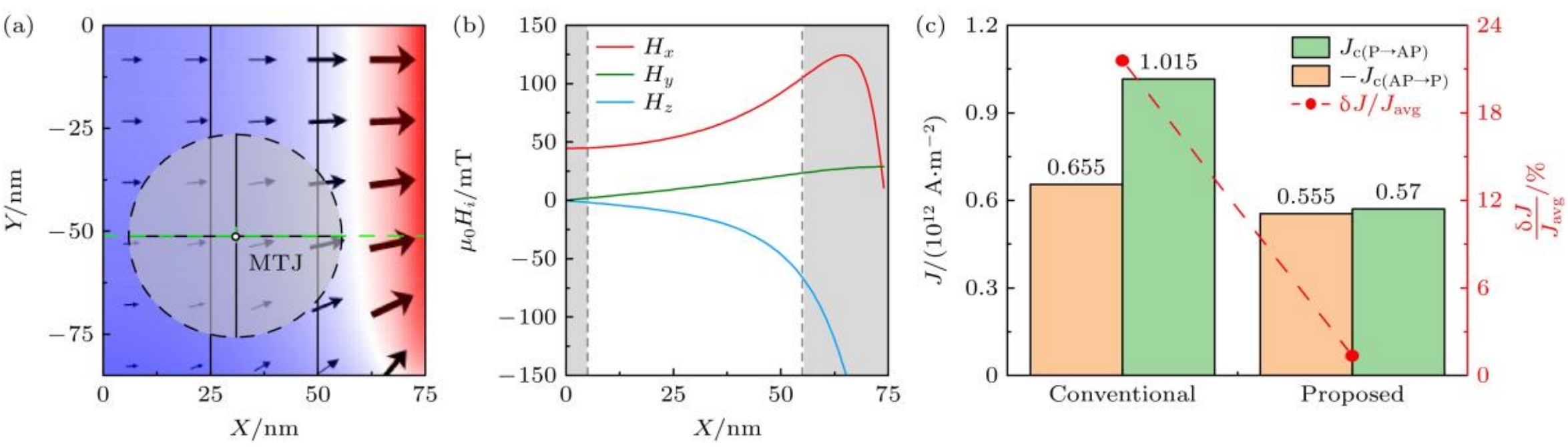

**Fig. 9.** Optimized local magnetic field distribution and performance comparison between the proposed and conventional architectures: (a) Magnetic field distribution around the MTJ when the center is offset to (30, −52); (b) distribution profiles of magnetic field components along the X-axis at Y = −52 nm; (c) comparison of critical switching current density and bias ratio between the proposed magnetic filling architecture (Proposed) at the optimized position (30, −52) and the conventional architecture (Conventional).

The proposed architecture can also be applied to devices with in-plane magnetic anisotropy (IMA). Because PMA and IMA structures differ substantially in their switching dynamics and magnetic-field response, the IMA implementation requires its own geometric parameters and magnetic filling configuration. To assess the generality of the concept, we designed and simulated an IMA variant; Appendix B gives the parameter set and bias-suppression results.

Micromagnetic simulations thus show that the magnetic filling architecture strongly suppresses write-current bias in both PMA and IMA structures, reducing the bias ratio to approximately 1% while maintaining deterministic switching without an external auxiliary field.

## 4 Scaling Performance

As semiconductor manufacturing processes evolve toward N14, N10, and even more advanced N5 nodes, spin-orbit torque magnetic random-access memory (SOT-MRAM) cells face stringent dimensional-scaling requirements. For instance, at the N14 process node, the MTJ diameter may be scaled down to 21 nm or smaller[36,37]. In this context, the excellent thermal stability of PMA structures, along with their advantages in high device integration density and low power consumption, makes them a promising platform for high-density spintronic devices[38]. The micromagnetic simulations in the preceding sections showed effective bias suppression in both PMA and IMA structures at larger dimensions. However, future extreme scaling scenarios impose higher requirements on architectural robustness. We therefore focus on PMA structures and evaluate whether effective write-bias suppression is retained as the key dimensions are scaled, thereby assessing the architecture's scalability.

To achieve precise bias compensation under scaling conditions, we first simulated the mechanism by which the geometric parameters of the magnetic filling layer regulate the local stray magnetic field distribution. As shown in Figure 10, we established a coordinate system whose origin is the geometric center of the upper surface of the heavy-metal layer and selected the coordinate (26, 0) as the characteristic monitoring point. Simulation results indicate that with a fixed magnetic-layer spacing T = 150 nm, the local magnetic field strength at the characteristic point increases significantly with the thickness H of the inserted magnetic layer, as shown in Figure 10(b). Conversely, when the magnetic layer thickness is fixed at H = 19 nm, the magnetic field strength increases nonlinearly as the magnetic-layer spacing T decreases, as shown in Figure 10(c). These results demonstrate that H and T constitute two independent and efficient degrees of freedom for tuning. This implies that as device scaling reduces T, one can reduce H to compensate for excessive stray fields. This approach maintains the appropriate compensating-field strength in the target region across different process nodes.

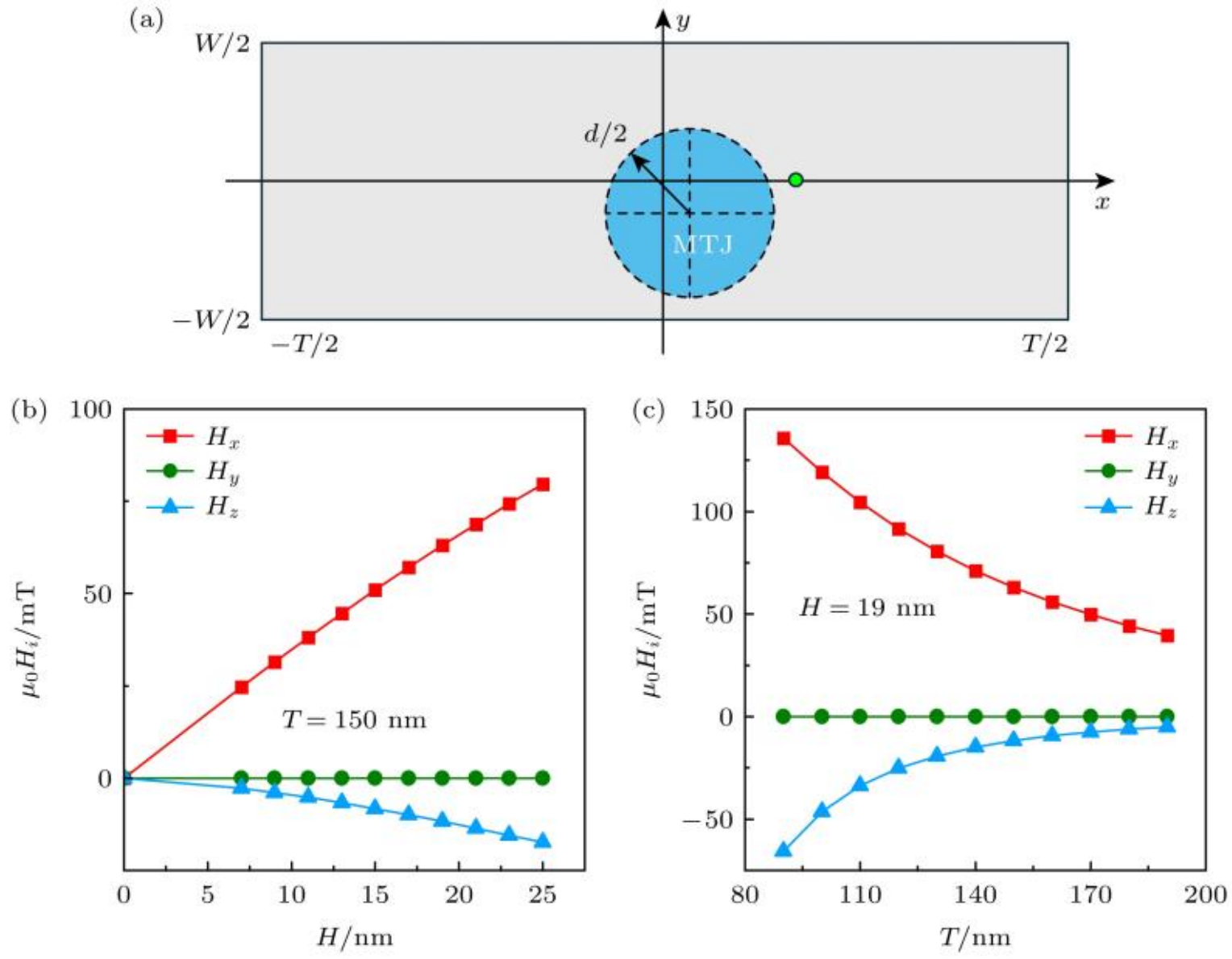

**Fig. 10.** Influence of magnetic layer thickness and spacing on the local magnetic field distribution: (a) Schematic of the relative coordinate system established on the free layer plane (with the origin at the geometric center of the heavy metal layer and a unit length of 1 nm); (b) variation of magnetic field components at (26, 0) as a function of magnetic layer thickness H with a fixed spacing T = 150 nm; (c) variation of magnetic field components at (26, 0) as a function of magnetic layer spacing T with a fixed thickness H = 19 nm.

Based on these trends, we proportionally scaled the key dimensions of the PMA cell: the MTJ diameter d, magnetic-layer spacing T, and heavy-metal length L and width W. Figure 11(a) shows the geometry. The simulations covered four scale factors: 80%, 60%, 40%, and 20%, corresponding to MTJ diameters of 40 nm, 30 nm, 20 nm, and 10 nm, respectively. To accommodate the field-strength changes at each scale factor, the magnetic filling layer thickness H was optimized to 15 nm, 11 nm, 9 nm, and 7 nm, respectively. Figures 11(b)–(e) present the performance comparison between the proposed architecture and the conventional architecture across different scale factors. Simulation results demonstrate that the proposed architecture exhibits excellent scaling robustness. At the 80% scale factor, placing the MTJ center at (23, −45) reduced $\delta J/J_{avg}$ from 15.7% in the conventional architecture to 0%. At the 60% scale factor, the bias ratio of the MTJ at (3, 0) decreased significantly from 11.0% to 0.6%. As the dimensions were further reduced to the 40% scale factor, selecting the optimized position (5, −15) suppressed the bias ratio from 6.8% to 0.1%. Notably, even under the extreme scaling condition of 20% (corresponding to an MTJ diameter of only 10 nm), the architecture significantly reduced the bias ratio from 3.4% in the conventional architecture to 0.1% when the MTJ center was located at (2, −2). These results show that the proposed architecture effectively mitigates the negative impact of the equivalent bias field $H_s$ at scaled dimensions. It demonstrates strong scalability, providing a viable technical pathway for the continuous evolution of SOT-MRAM toward high density and low power consumption.

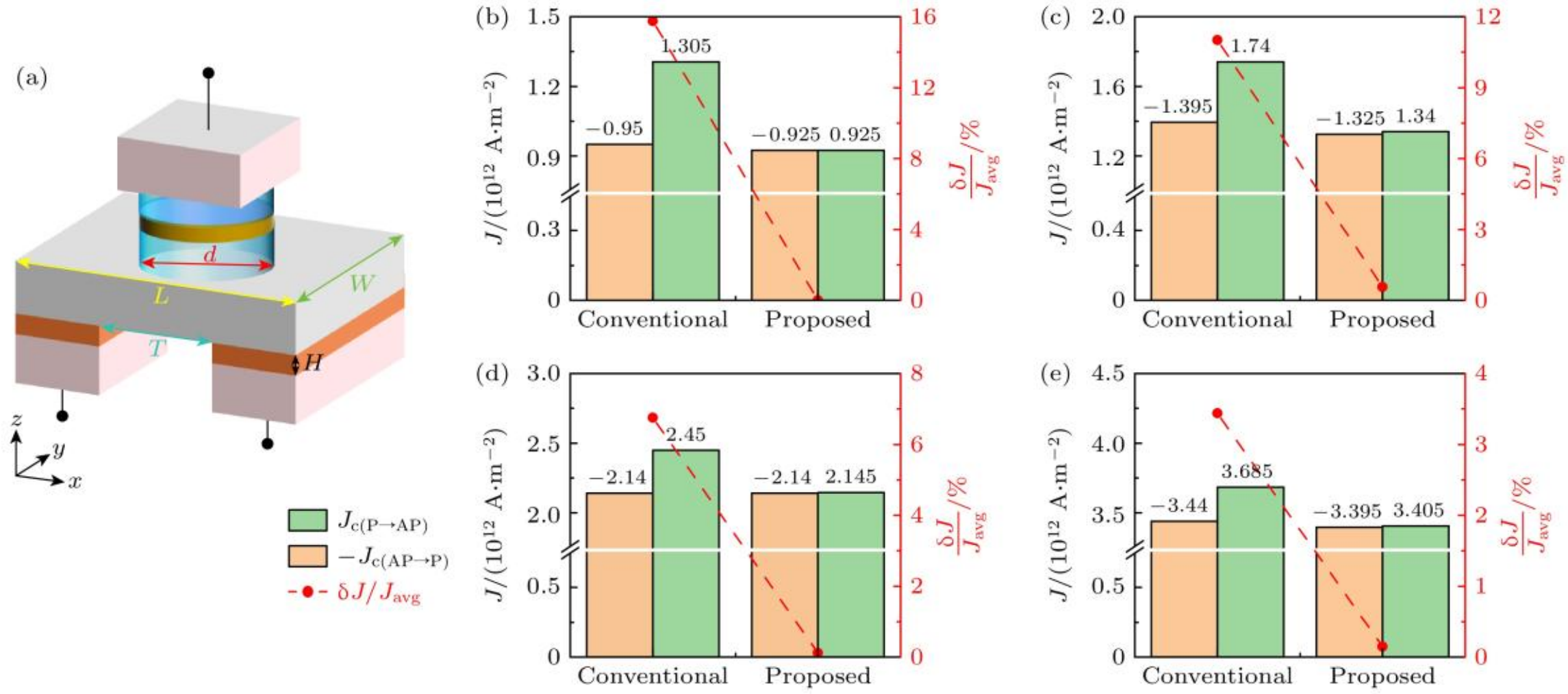

**Fig. 11.** Proportional scaling of the PMA structure and comparison of its performance: (a) device geometry used for the scaling study; (b)-(e) critical switching current densities and bias ratios of the conventional and proposed magnetic filling architectures at identical dimensions, scaled to (b) 80%, (c) 60%, (d) 40%, and (e) 20%.

## 5 Conclusion

This study addresses critical bottlenecks in spin-orbit torque magnetic random-access memory (SOT-MRAM), specifically the write-current asymmetry caused by bias magnetic fields and the dependence on external auxiliary fields. We propose a novel cell architecture with a magnetic filling layer that is highly compatible with back-end-of-line (BEOL) processes. By combining micromagnetic simulations with theoretical analysis, we establish the physical mechanism for using localized stray magnetic fields to compensate for reference layer stray fields and interlayer coupling effects. We systematically verify the effectiveness of this architecture across different magnetic anisotropy systems. Simulation results demonstrate that tailoring the magnetization direction and geometric parameters of the magnetic filling layer enables deterministic switching without external field assistance. For PMA structures, the write-current bias ratio decreases significantly from 21.6% in the conventional architecture to 1.3%. For in-plane magnetic anisotropy (IMA) structures with high-speed field-free writing potential, specifically the $\varphi = 30°$ configuration, the architecture suppresses the bias ratio from 19.8% to −0.2%. These results confirm the generality and effectiveness of the concept across the two anisotropy configurations considered.

Building on this mechanism, we further evaluated the architecture at advanced semiconductor technology nodes. To address scaling at N14 and beyond, the thickness of the magnetic filling layer and the spacing between the two filling sections were reduced in concert. Even when all device dimensions were scaled to 20% of their original values (an MTJ diameter of 10 nm), the simulated bias ratio remained at 0.1%. Unlike conventional compensation schemes that add magnetic layers within the MTJ stack and are highly sensitive to interfacial coupling and film-thickness control, the proposed BEOL magnetic filling scheme leaves the core MTJ stack unchanged. This design may therefore reduce process complexity and cost while retaining compatibility with advanced technology nodes.

Although the simulations support the theoretical feasibility and performance advantages of the architecture, substantial technical barriers remain between the idealized model and practical implementation. Future work will focus on experimental validation using fabricated devices, the controllability and uniformity of magnetic filling at wafer scale, and possible magnetic coupling between filling layers in adjacent cells within large, dense arrays. Such studies are needed to establish a complete engineering evidence base. In summary, the proposed magnetic filling architecture addresses both write-current asymmetry and external-field dependence at the device-physics level and shows promising compatibility with advanced semiconductor technology nodes. It provides a potential route toward high-density, low-power, and highly reliable next-generation magnetic memory arrays.

## Appendix A: Power-Consumption Benefit of the Magnetic Filling Layer

The scaling relationship for power consumption is $P \propto I^2 R$. This paper has shown that when a current bias is present, the critical current density can be expressed as $|J_{AP\to P}| = J_{avg}-|\delta J|$ and $|J_{P\to AP}| = J_{avg}+|\delta J|$. We define the bias ratio as $\delta J/J_{avg}$. In an engineering implementation, if a "unified write current" is adopted, the magnitude of the write current must be at least $|J| = J_{avg}+|\delta J| = J_{avg}(1+|\delta J/J_{avg}|)$. Therefore, assuming other conditions remain constant and modeling the write path as a resistance R, the write power consumption satisfies the following relationship with the bias ratio:

$$P \propto (1 + |\delta J/J_{\mathrm{avg}}|)^2 R. \tag{A1}$$

Consequently, reducing the bias ratio lowers the required unified-write-current power according to the squared factor in Eq. (A1). This finding aligns with the conclusion that lowering the bias ratio directly decreases power dissipation. In practical simulations, incorporating magnetic filling structures reduces $J_{avg}$. However, to directly evaluate the power benefits of the proposed architecture, we neglect the power reduction from decreased $J_{avg}$. We consider only the power savings attributable to the reduced bias ratio.

We estimate power consumption using the geometric parameters of the proposed PMA cell. In the cell used to verify the compensation effect, the magnetic filling layer thickness (equivalent to the current path length within the magnetic segment) is H = 19 nm. Under optimized structural conditions, the bias ratio decreases significantly from 21.6% in conventional architectures to 1.3%. Therefore, when evaluating solely based on a unified write current, the power scaling factor resulting from bias suppression is

$$\frac{P_{\mathrm{new}}}{P_{\mathrm{old}}}\Big|_R = \left(\frac{1+0.013}{1+0.216}\right)^2 \approx 0.69, \tag{A2}$$

Thus, improving write-current symmetry alone corresponds to an approximately 30% reduction in power consumption.

We estimate the additional power associated with the resistivity of the magnetic filling layer using conservative limiting values. The resistance of a VIA segment is R = ρH/A, where A is its effective conductive cross-sectional area. For a conservative worst-case estimate, we take a relatively small cross section A = $(50\ \mathrm{nm})^2$, consistent with the cell dimensions. For the conventional nonmagnetic VIA, we use tungsten (W), with $\rho_W = 11.5 \times 10^{-8}\ \Omega\cdot\mathrm{m}$, as an optimistic lower bound[39]. For the CoFeB magnetic filling layer, Ref. [40] reports $\rho_{CoFeB} \approx 165 \times 10^{-8}\ \Omega\cdot\mathrm{m}$ at nanoscale thickness; we round this value up to $\rho_m = 200 \times 10^{-8}\ \Omega\cdot\mathrm{m}$ as a conservative upper bound. The resulting resistance of one VIA is

$$R_{\mathrm{VIA,W}} = \rho_{\mathrm{W}} H/A \approx 0.87\Omega, R_{\mathrm{VIA,m}} = \rho_{\mathrm{m}} H/A \approx 15.2\Omega. \tag{A3}$$

Because the architecture contains magnetic filling in two symmetric VIAs, the upper bound of the additional series resistance for the two VIAs is approximately

$$\Delta R_{\mathrm{2VIA}} \approx 2(R_{\mathrm{VIA,m}} - R_{\mathrm{VIA,W}}) \approx 28.7\Omega. \tag{A4}$$

We compare the power increase from added resistance with the current reduction from bias suppression. The total write-path resistance is $R_{tot} = R_{HM}+R_{2VIA}$, where $R_{HM}$ denotes the heavy-metal-channel resistance. Using W as a representative spin-orbit material, Ref. [41] reports a fitted W-layer resistivity of approximately $130 \times 10^{-8}$ Ω·m for W/CoFeB/MgO. Using $\rho_{HM} = 130\times10^{-8}$ Ω·m and a representative heavy-metal thickness t = 4 nm, we obtain:

$$R_{\mathrm{HM}} = \rho_{\mathrm{HM}} \frac{L}{Wt} \approx 803\Omega. \tag{A5}$$

Using these values, replacing the nonmagnetic VIA metal with the magnetic filling layer gives a total resistance ratio of

$$\frac{R_{\mathrm{tot,new}}}{R_{\mathrm{tot,old}}} \approx \frac{R_{\mathrm{HM}}+R_{\mathrm{2VIA,Cu}}+\Delta R_{\mathrm{2VIA}}}{R_{\mathrm{HM}}+R_{\mathrm{2VIA,Cu}}} \approx 1.036. \tag{A6}$$

Considering both current and resistance factors, the power consumption ratio under the unified write current mode is

$$\frac{P_{\mathrm{new}}}{P_{\mathrm{old}}} \approx \left(\frac{1+0.013}{1+0.216}\right)^2 \times 1.036 \approx 0.71. \tag{A7}$$

Consequently, with W used for both the VIA metal and the heavy-metal channel, the proposed scheme gives an estimated net power reduction of approximately 29%. Advanced BEOL interconnect materials vary with technology node and reliability requirements, with Cu and Co among the available options[42]. To test sensitivity to interconnect material, we also evaluated a low-resistance limit using ideal Cu for the interconnect and Pt for the heavy-metal layer. The estimated net power reduction remains approximately 16%. Thus, for both parameter sets considered, the added series resistance is insufficient to offset the power benefit of bias suppression.

## Appendix B: Bias Compensation by the Magnetic Filling Layer in the IMA Structure

In the IMA structure, sub-nanosecond writing is achievable at φ = 0°, but it requires an external magnetic field to assist switching of the free layer. At φ = 90°, field-free switching is possible, but the dynamics exhibit an incubation delay similar to that in STT-driven switching. When the in-plane easy axis forms an angle of 0° < φ < 90° (here φ = 30°), high-speed, field-free writing can be achieved. Therefore, IMA structures with 0° < φ < 90° are generally considered more promising for practical applications. On this basis, we simulated SOT-MRAM with the proposed architecture at φ = 30° and compared it with the conventional architecture.

Based on previous studies and the present simulation parameters, we set the equivalent bias field of the IMA structure to $\mu_0 H_s = 8$ mT, using $H_c/H_s \approx 10/1$[9,43]. The IMA device uses the same overall geometry as the PMA device, except that its MTJ free layer is an elliptical cylinder with major axis a = 50 nm, minor axis b = 25 nm, and thickness 1.5 nm. The magnetization direction of the magnetic filling layer in the VIA is along the +z direction, as shown in Fig. B1.

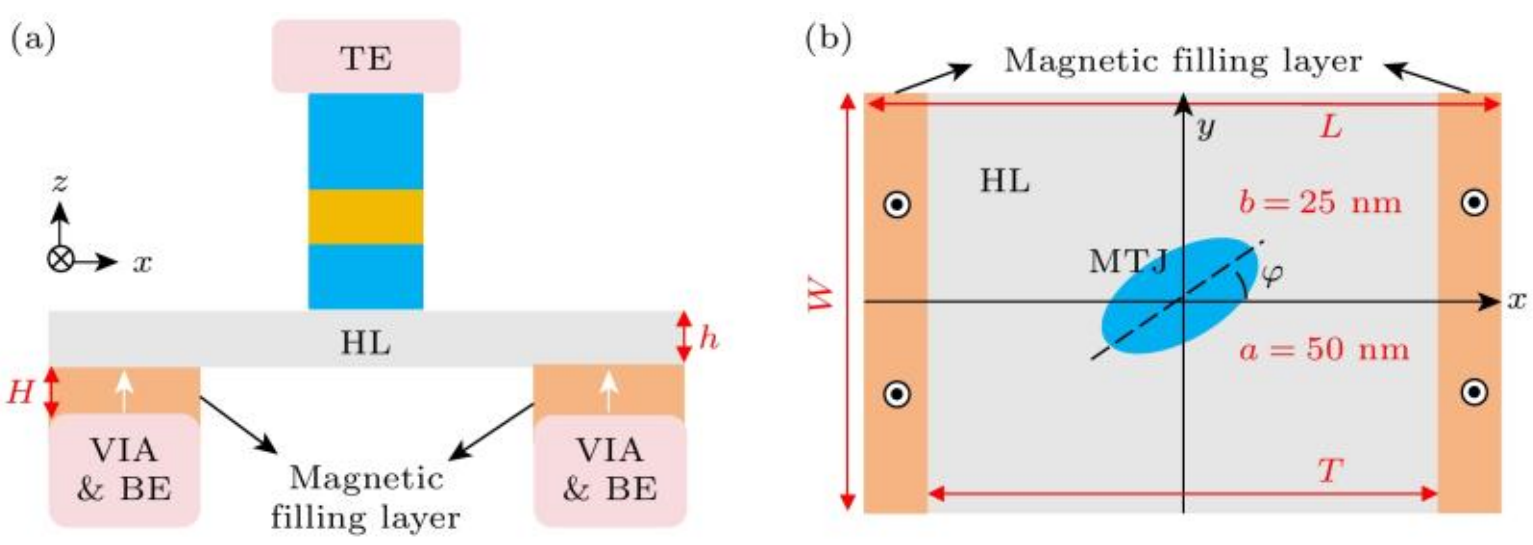

**Fig. B1.** Schematic of an IMA SOT-MRAM cell containing magnetic filling: (a) front view; (b) top view.

Similar to the analysis of the PMA structure, we characterized the magnetic field distribution within the gray rectangular region of the free layer plane in Fig. B1(b), as shown in Fig. B2(a). We extracted the distribution curves of magnetic field components along various directions as a function of position along the green dashed line in Fig. B2(a), as presented in Fig. B2(b). Based on the theoretical analysis in Section 2, the fourth quadrant provides local magnetic field components suitable for compensating the bias induced by $H_s$ in the IMA structure. Therefore, we conducted a refined search in the positive x-axis region and selected the MTJ center coordinate (25, −25) as a typical optimization position. We compared the performance of the proposed magnetic filling architecture with that of the conventional architecture at this position. The results indicate that the bias ratio decreased significantly from 19.8% in the conventional architecture to −0.2%, effectively eliminating the negative impact of the equivalent bias field $H_s$, as shown in Fig. B2(c).

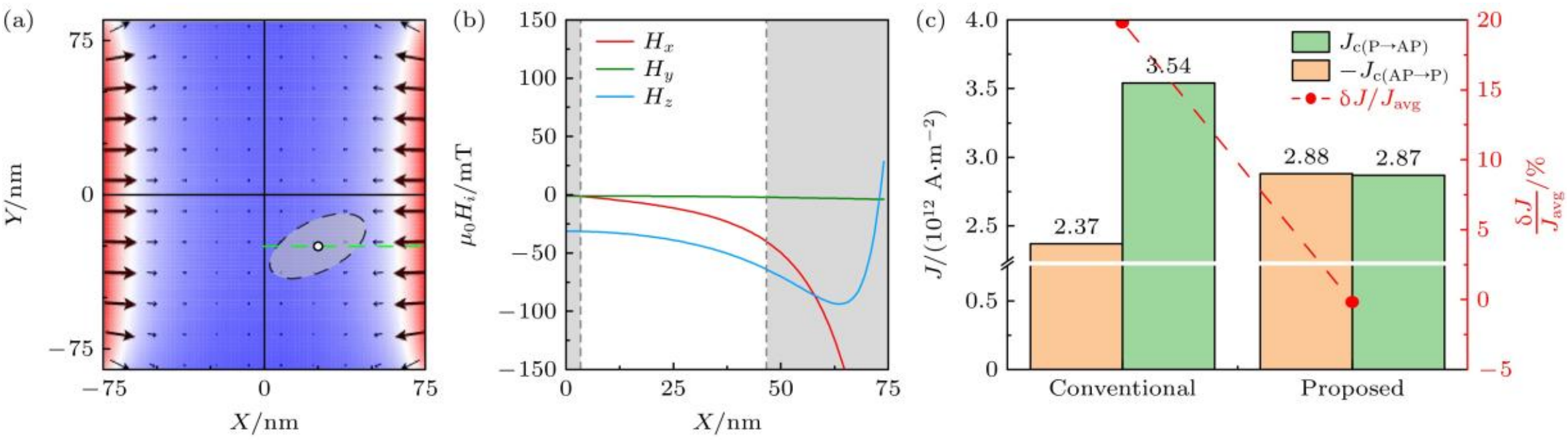

**Fig. B2.** Local magnetic-field distribution and position-dependent compensation of switching-current bias: (a) magnetic-field distribution in the gray rectangular region in Fig. B1(b); (b) magnetic-field components along the green dashed line in Fig. B2(a); (c) critical switching current densities and bias ratios of the proposed magnetic filling architecture at the optimized position (25, −25) and of the conventional architecture.

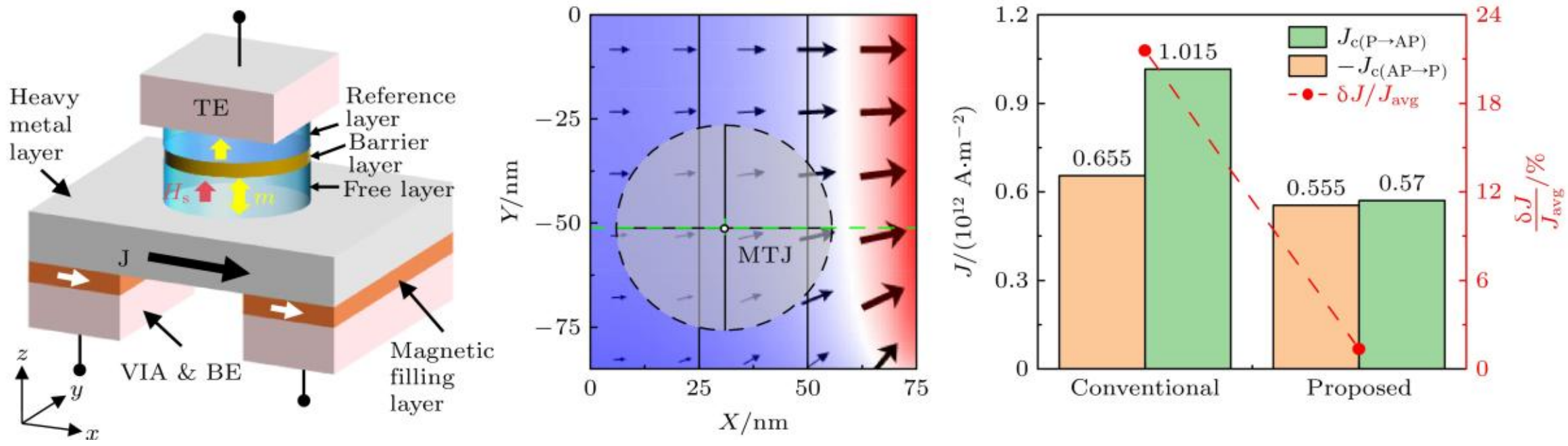